\documentclass[preprint,12pt]{elsarticle}
\usepackage{amssymb}
\usepackage{array,graphicx}
\usepackage[nointegrals]{wasysym} 
\usepackage[normalem]{ulem}
\usepackage{makecell}
\usepackage{soul}
\usepackage{rotating}
\usepackage{epstopdf}
\usepackage{enumitem}
\usepackage{siunitx}
\usepackage{svg}
\usepackage{tablefootnote}
\usepackage{balance}
\usepackage{xurl}
\usepackage{hyperref}
\usepackage{xcolor}
\usepackage[most]{tcolorbox}
\usepackage{float}
\usepackage{multirow}
\usepackage{tikz}
\usepackage{tikz,pgfplots}
\usepackage{pgfplots}
\usepgfplotslibrary{groupplots}
\usepgfplotslibrary{statistics}
\usepackage[titlenumbered,linesnumbered,ruled,vlined]{algorithm2e}
\usepackage{bm}
\usepackage{framed}
\usepackage{algpseudocode}
\usepackage{graphicx}
\usepackage{url}
\usepackage{hyperref}
\usepackage{dblfloatfix}
\usepackage{array}
\usepackage{enumitem}
\usepackage{multicol}
\usepackage{multirow}
\usepackage{booktabs}
\usepackage{amsmath}
\usepackage{mathtools}
\usepackage{float}

\NewDocumentCommand\mm{g}{%
  \IfNoValueF{#1}{{\color{blue} \textbf{(MM: #1)}}}%
  \IfNoValueT{#1}{{\color{blue} \textbf{(MM)}}}%
}
\NewDocumentCommand\hg{g}{%
  \IfNoValueF{#1}{{\color{orange} \textbf{(HG: #1)}}}%
  \IfNoValueT{#1}{{\color{orange} \textbf{(HG)}}}%
}
\NewDocumentCommand\er{g}{%
  \IfNoValueF{#1}{{\color{purple} \textbf{(HG: #1)}}}%
  \IfNoValueT{#1}{{\color{purple} \textbf{(HG)}}}%
}

\newcommand{\subhead}[1]{\vspace {1pt}\noindent{\textbf{#1.}}}

\newcolumntype{C}{>{\centering\arraybackslash}p{1.5cm}}
\newcommand{\name}{Blindfold}

\journal{Computers and Security}

\begin{document}

\begin{frontmatter}

\title{Blindfold: \\Keeping Private Keys in PKIs and CDNs out of Sight}

\author[inst1,inst2]{Hisham Galal\corref{cor1}}
\ead{hisham_sg@aun.edu.eg}
\author[inst1]{Mohammad Mannan}
\ead{m.mannan@concordia.ca}
\author[inst1]{Amr Youssef}
\ead{youssef@ciise.concordia.ca}

\affiliation[inst1]{organization={Concordia Institute for Information Systems Engineering, Concordia University},
            city={Montreal},
            state={Quebec},
            country={Canada}}
\affiliation[inst2]{organization={Faculty of Computers and Information, Assiut University},
            country={Egypt}}
\cortext[cor1]{Corresponding author}

\begin{abstract}
Public key infrastructure (PKI) is a certificate-based technology that helps in authenticating systems identities. HTTPS/TLS relies mainly on PKI to minimize fraud over the Internet. Nowadays, websites utilize CDNs to improve user experience, performance, and resilience against cyber attacks. However, combining HTTPS/TLS with CDNs has raised new security challenges. In any PKI system, keeping private keys \emph{private} is of utmost importance. However, it has become the \emph{norm} for CDN-powered websites to violate that fundamental assumption. Several solutions have been proposed to make HTTPS CDN-friendly. However, protection of private keys from the very instance of generation;  and how they can be made secure against exposure by malicious (CDN) administrators and malware remain unexplored. We utilize trusted execution environments to protect private keys by \emph{never} exposing them to human operators or untrusted software. We design \name\ to protect private keys in HTTPS/TLS infrastructures, including CAs, website on-premise servers, and CDNs. We implemented a prototype to assess \name's performance and performed several experiments on both the micro and macro levels. We found that \name\ slightly outperforms SoftHSM in key generation by 1\% while lagging by 0.01\% for certificate issuance operations.
\end{abstract}

\begin{keyword}
Key management \sep Trusted Execution Environment \sep PKCS\#11
\end{keyword}

\end{frontmatter}

\section{Introduction}
Many public-key cryptography protocols can be proven secure under reasonable assumptions. However, in the majority of these protocols, security guarantees hold only as long as private keys remain \emph{private}. In practice, fulfilling that simple assumption is more complicated than one might expect, given the many incidents where private keys are exposed accidentally or deliberately via an attack. The consequences of exposed private keys are highly severe, especially in protocols utilizing digital signatures, as attackers can launch impersonation attacks and generate fraudulent signatures. 

The impact of exposed private keys can be clearly seen across different domains~\cite{5772960, codecov, prins2011diginotar, ze, symantec, cryptoHacks2,cryptoHacks1}. In the context of web PKI, an attacker managed to steal private keys from Certification Authority (CA) DigiNotar~\cite{prins2011diginotar} and issued over 500 fraudulent certificates for top Internet companies like Google, Mozilla, and Skype. Similarly, in code signing, Stuxnet~\cite{5772960} abused exposed private keys for hardware driver certificates from Realtek and JMicron~\cite{falliere2011w32}, and succeeded in code signing its malicious drivers. Operating systems and browsers revoked all exposed private keys; however, too late after causing devastating consequences to the extent that some companies declared bankruptcy~\cite{ze}. These incidents are consequences of direct attack, and they can be mitigated to some extent by following better security practices and policies. Unfortunately, in some domains, such as cryptocurrency, the damage of exposed private keys is final and cannot be mitigated.

Interestingly, as one might not expect, there are other circumstances where private keys owners voluntarily share their keys with third parties, which can be essentially considered as exposure of private keys from the PKI stand-point. This problem has been first noted by Liang et al.~\cite{6956557} has been studied at a large scale by Cangialosi et al.~\cite{10.1145/2976749.2978301}, highlighting the widespread sharing of website private keys with CDNs and cloud/hosting service providers. In some cases, third parties play an active role in managing their customers’ keys, and even several providers afford to apply for certification on behalf of their customers. 

Many solutions~\cite{10.1145/3127479.3127482,247664,8567660, 9343007} have been proposed over the years to make HTTPS CDN-friendly without exposing private keys. However, they do not address how such keys are protected from the very beginning of them being generated, provisioned, and copied for backup. In this work, we take a fundamentally different approach to protecting private keys by never exposing them to human owners, administrators, or unauthorized software --- trusting only hardware security features offered by contemporary CPUs to safeguard these keys. Different PKI-based systems can utilize our generic approach for protecting private keys against exposure over their entire life-cycle. 

As a concrete realization of this approach, we design \name\footnote{\name\ is an X-Men hero born without eyes.} that keeps private keys out of sight by maintaining them within a particular CPU protected environment. We show how \name\ can protect HTTPS certificates' private keys on-premises web servers, remotely on CDN nodes, and even safeguard CAs' private keys for root and intermediate certificates. In \name,  all parties \emph{attest} each other, i.e., check for the presence of expected software and hardware Trusted Execution Environments (TEEs), e.g., before issuing a certificate, provisioning a private key at an owner/CDN-controlled machine, and securely backing up private keys. A key objective of \name's design is to introduce minimal configuration changes to existing systems. Fortunately, modern web servers and CA systems such as Apache, NGINX, and Boulder~\cite{boulder}, which is an ACME server, and even OpenSSL library support integration with the Public Key Cryptography Standard (PKCS\#11~\cite{standard2020pkcs}) interface. Hence, we design \name\ to adhere to the PKCS\#11 interface.

The contributions of this paper include:
\begin{enumerate}
	\item We design \name\ as a generic approach that prevents private keys exposure even against significant threats, including system operators and malware, by leveraging TEE capabilities in modern CPUs. 

	\item We build a concrete instance of \name\ to address the problem of certificates' private keys sharing by websites with CDNs and other hosting providers. 

	\item We solve the private keys exposure problem by enforcing \emph{attestation} to the whole generation of private keys by \name\ running on an authentic TEE platform. 

	\item We carry out several experiments on a prototype on the micro and macro levels to measure the performance of \name.
\end{enumerate}

\section{Key Sharing, Exposure Incidents, and Related Work}
\subsection{Private Key Exposure and Sharing Practices}
\subhead{CA incidents}
In the setting of certificate issuance, both parties (i.e., a CA and a website) have private keys albeit for different purposes: (i) CA private key (i.e., root and intermediate keys) for signing certificates, and (ii) website private key for authenticating its web domain during TLS handshake with browsers. Essentially, browsers must revoke millions of certificates if the CA private key is exposed. Accordingly, this will create havoc as millions of websites will suddenly find their certificates are no longer trusted, and browsers will warn users when they try visiting these websites. To get a rough idea about the severe consequences of such an event, Let's Encrypt has recently revoked over three million certificates~\cite{revoke} due to a software \emph{bug}, and gave the affected websites 24 hours to resolve their issues. Indeed, this number could be much higher if the revocation was due to Let's Encrypt private key compromise. 

Several CAs in the past have been found deviating from the guidelines and requirements regarding certificate issuance and management of private keys as set by the Certification Authority Browser (CAB) Forum~\cite{cab}. A comprehensive study by Serrano et al.~\cite{serrano2019complete} revealed notable incidents of private key exposure, which include: eight cases by Comodo, five by WoSign, four by Symantec and VeriSign, two by DigiCert, and one by DigiNotar, India CCA, Let's Encrypt, StartCom and Thawte. In most cases, operating systems and browsers distrusted CAs after publicly published incidents. Even more, several companies declared bankruptcy. Besides, websites certificates issued by those CAs had their business disrupted. Indeed, there is no guarantee that CAs publish all internal incidents as such disclosure can negatively impact their business. Hence, one cannot take for granted that all CAs protect their private keys according to the CAB Forum guidelines. Therefore, one of the main objectives for \name\ is to protect private keys while also providing attestation that can be publicly verified (e.g., by the websites at the time of certificate issuance). 

\subhead{Key-leakage incidents}
One more reason to motivate the need for enforcing the protection of private keys against exposure is the shocking number of key leakage incidents. A recent report~\cite{meli2019bad} in 2019 shows that thousands of cryptographic keys are leaked on GitHub every day. In many cases, the private keys were included in the source code repositories~\cite{source}, bundled within the binary resources of the software~\cite{dlink}, or even published on blog posts by manufacturers~\cite{adobe}. For example, a user who downloaded the updated firmware from the D-Link~\cite{dlink} found not only the private keys but also the passphrase for the code signing certificate bundled in the firmware code. This simple mistake of packaging the wrong files in the final binary leads to potentially serious consequences. Typically, malware authors use leaked certificate keys to sign their malware, bypassing signature checks by operating systems~\cite{5772960,kim2017certified}.

\subhead{Heartbleed vulnerability}
The Heartbleed vulnerability ~\cite{10.1145/2663716.2663755,zhang2014analysis} is a bug in the OpenSSL library that was publicly disclosed in 2014. It allows attackers to scan servers’ memory, which in turn could potentially reveal the servers’ plaintext private keys (even if the private keys remained encrypted, e.g., by a passphrase on-disk following PKCS\#8~\cite{standard2020pkcs}). Consequently, an essential objective of \name\ is to protect private keys in a secure memory environment that is not vulnerable to attackers or malicious operators.

\subhead{Key sharing practices with CDNs} 
Websites increasingly use third-party providers such as CDNs and hosting providers to host HTTPS content, sharing their certificates' private keys. For example, Cangialosi et al.~\cite{10.1145/2976749.2978301} studied at a large scale the key sharing problem taking place in practice. In particular, websites share their keys with hosting providers in two ways. First, a website may explicitly turn over control of its private key by simply uploading it to the hosting provider. Second, a hosting provider may access the website's private key via a web service. In both cases, the website is not only trusting the CDN or hosting provider not to abuse the keys but also to prevent both external and internal attackers from accessing the keys.

The severity of a private key exposure is ranked according to the role of its owner as follows:
\begin{enumerate}
    \item CA: This is the most devastating case where an adversary breaching a CA will have access to the private keys for the CA certificate. Thus, the adversary can issue rogue domain certificates that browsers trust.
    \item CDN: The adversary can access the private keys shared by websites hosting content on the CDN. Thus, the adversary can snoop users' information and impersonate affected websites. Furthermore, the adversary can also impersonate any website with its domain name listed in cruise-liner~\cite{10.1145/2976749.2978301} certificates controlled by the CDN.
    \item Web server: This attack occurs if an adversary compromises an origin web server that might be hosting a few websites. Thus, it has similar consequences as a compromise on a CDN, yet with much fewer affected websites. 
\end{enumerate}

The widely considered solution to mitigate these breaches is to revoke the compromised certificates; however, the number can scale from a few certificates to millions in the case of a breach on a web server or a CA, respectively. Indeed, such a mitigation solution is expensive in terms of time and resources besides financial loss due to business disruption. Therefore, it is paramount to prevent private key exposure rather than relying only on mitigation.

\subsection{Related Work}
In Table~\ref{tb.related}, we list the features provided by \name\ and other constructions used in production such as Fortanix~\cite{fort}, 
Cloud HSM~\cite{awshsm,ibmhsm}, and Keyless SSL~\cite{keyless}.

\begin{table}[ht]
\centering
\caption{Comparing features of \name\ against other constructions. Solid, partial, and empty circles denote full, partial, and lack of  feature support, respectively.}
\label{tb.related}
\begin{tabular}{@{}lcccc@{}}
	\toprule
	 &\name\ & Fortnix& Cloud HSM& Keyless SSL\\
	\midrule
Remote Attestation & \CIRCLE& \CIRCLE& \Circle & \Circle \\
PKCS\#11 compatible &\CIRCLE &\Circle &\CIRCLE &\Circle\\
Scalability &\CIRCLE &\RIGHTcircle &\RIGHTcircle &\Circle\\
No extra round-trip &\CIRCLE &\Circle &\RIGHTcircle & \Circle\\
Self-custody of keys &\CIRCLE &\RIGHTcircle & \Circle & \CIRCLE\\
\bottomrule	
\end{tabular}
\end{table}
		
Fortanix~\cite{fort} claims their product Self-Defending Key Management Service (SDKMS)  to be the first cloud service secured with Intel SGX. SDKMS enables clients to generate cryptographic keys and certificates securely. However, it is a centralized service hosted in the cloud with limited scalability. Moreover, for an HTTPS/TLS request, there is an extra round-trip to access the private keys. 

Amazon AWS Cloud HSM~\cite{awshsm} and IBM Cloud HSM~\cite{ibmhsm} provide services to protect cryptographic keys and securely perform operations. They enable clients to generate cryptographic keys on the Cloud HSM via standard interfaces such as PKCS\#11. In addition, cloud HSM provides the minimum level of attestations to generating private keys. One can verify these attestations by verifying the certificate chain on the generated key leading to the root certificate to the HSM manufacturer. However, with Intel SGX attestation, \name\ can achieve finer control on what needs to be attested on arbitrary state and functionality.

\par Furthermore, although HSMs support the generation of non-extractable private keys, these keys cannot be copied individually to a destination HSM. The only solution is to clone the entire HSM, which copies all cryptographic keys from the source HSM. It becomes a problem when a website wants to provide its non-extractable certificate private key to a CDN without cloning all keys.  

\par Additionally, the throughput of cryptographic requests is affected by the resource utilization of Cloud HSM and its network status since there is a limited number of HSM slots a cloud operator can offer. This problem is even worse when a CDN containing thousands of nodes sends requests to Cloud HSMs, introducing extra round-trips to serving HTTPS requests. Moreover, using Cloud HSM is quite expensive, e.g., AWS Cloud HSM and IBM Cloud HSM charge \$1,168 and \$1,250 per month, respectively (as of June 2021). Conversely, in \name\, the website controls what keys can be provisioned. Additionally, each CDN node can have its own \name\ instance; hence there is no congestion or extra round-trips. Moreover, Intel SGX is cheaper than a monthly subscription to Cloud HSM service on commodity hardware.

To ensure the protection and self-custody of a certificate's private key, CloudFlare offers \textit{Keyless SSL}~\cite{keyless}. Essentially, the CDN decomposes the TLS handshake protocol. Then, all operations related to the certificate's private key are forwarded to the customer's key server, which replies to the operations' results to the CDN. Hence, the CDN can complete the TLS handshake and establish session keys without controlling their customers' private keys. The main problem with this approach is that for each TLS handshake, the CDN must contact the client's key server. Therefore, it adds an extra round-trip latency which may degrade the overall performance.
On the contrary, \name\ can be installed in CDN nodes, thereby eliminating the need for extra round-trips.

\section{Background and Threat Model}
This section provides a brief overview of some necessary components of the current PKI ecosystem and Intel SGX environment used as a TEE in our design. We also discuss our threat model and assumptions about different entities in our design.

\subsection{X.509 Certificate}
X.509~\cite{myers1999internet} is the standard format for defining a digital certificate that binds a signature to a public key, a subject, an issuer, a validity period, and a set of extensions. The extensions define extra attributes or constraints on the use of the certificate. For a certificate to be valid, it has to meet the following requirements:
\begin{enumerate}
    \item The certificate chain ends with a self-signed certificate that the client considers trusted (e.g., a browser or an operating system).
    \item The attributes of the certificate chain have valid parameters (e.g., the validity period of a certificate has not expired).
    \item Neither the leaf certificate nor any chain certificate is revoked. 
\end{enumerate}

Web browsers are initially configured with a set of trusted CA certificates. When a user opens an HTTPS session to a website, the browser receives a domain certificate and an intermediate certificate chain from the web server. Then, it validates the domain certificate as well as checks whether the \texttt{Subject} field matches the web domain.

To obtain a certificate, an applicant creates a Certificate Signing Request (CSR) as defined by PKCS\#10~\cite{standard2020pkcs} and sends it to the CA. The CA will validate whether the applicant truly owns the specified web domain. Upon success, the CA will convert the CSR into X.509 certificate and returns it to the applicant.

\subsection{Intel Software Guard Extensions (SGX)}
Intel SGX~\cite{anati2013innovative} is a TEE technology released by Intel in 2015. It provides an isolated secure environment referred to as \emph{enclave} for code and data that need to be protected against violations of confidentiality and integrity. Note that an enclave can be statically disassembled; hence, it must not contain any hard-coded secrets. However, once it is loaded and running, the processor enforces the confidentiality and integrity of the enclave state. Therefore, an observer will have an opaque view of the enclave's state, including any generated secret.

\subhead{Sealing} Intel SGX enclave can securely generate cryptographic keys at the run-time. However, all the generated keys will be lost once the enclave is torn down (e.g., on application exit or power event). Therefore, Intel SGX provides the ability to cryptographically \emph{seal}~\cite{anati2013innovative} secrets to untrusted external storage in a secure way. Encryption is performed using a private \emph{Seal} key that is unique to that particular platform and enclave. The derivation of that key depends on the enclave identity. 

\par A developer can set the enclave's identity either to \emph{MRENCLAVE} or \emph{MRSIGNER}. The former is a cryptographic hash of the enclave's code, data, and other measurements as it goes through every step of the build and initialization process. Hence, MRENCLAVE uniquely identifies any particular enclave. Thereby, only instances from the same enclave can decrypt the sealed data. The latter is provided by an authority that signs the enclave before its distribution. Thus, different enclaves signed by the same authority can derive the exact seal key.

\subhead{Remote attestation} Intel SGX provides the ability to cryptographically \emph{attest} that a particular enclave is running on an authentic Intel SGX platform. The attestation process starts with an application requesting its enclave to generate a \texttt{report}. The enclave generates the \texttt{report} and authenticates it by a platform-specific hardware key. The \texttt{report} contains enclave-specific information, notably, the measurement \emph{MRENCLAVE} and an auxiliary data \texttt{report\_data} field.  The main purpose of the \texttt{report\_data} field is to bind a piece of data (e.g., a public key of a private key generated exclusively by the enclave) to the \texttt{report}.

Next, the application transmits the \texttt{report} to an architectural enclave known as \emph{Quote Enclave} (QE) running on the same platform. After verifying the authenticity of the \texttt{report} using the same platform-specific hardware key, QE signs the \texttt{report} by the attestation key and returns a \texttt{quote} (i.e., a signed \texttt{report}) to the calling application, which is eventually transmitted to the remote party. Note that, besides attesting to authentic Intel SGX platforms, a valid quote also implies the authenticity of \texttt{report\_data}. For instance, this allows two enclaves to establish a secure channel after quotes verification by setting their ephemeral ECDH public keys in \texttt{report\_data} in their corresponding quotes. 

Currently, Intel SGX supports two models of remote attestations: EPID~\cite{johnson2016intel} and ECDSA~\cite{scarlata2018supporting}. In EPID attestation, the main focus is to preserve the attestor's privacy by utilizing an EPID group signature scheme. Hence, an entity with the group's public key can verify a quote without learning which group member (i.e., processor) has signed it. Intel exclusively maintains the group public key in its Attestation Service (IAS). Thereby, one must consult IAS to verify a quote, which replies with an attestation verification report that confirms or denies the quote's authenticity. 

Conversely, the ECDSA attestation sacrifices the attestor's privacy since it is mainly used to attest to platforms within the same organization. More importantly, it does not require communication with IAS. Therefore, ECDSA attestation is more convenient when (i) attestors' identities are already known in advance or (ii) when external communication with IAS is restricted. 
 \begin{figure*}[htb!]
    \centering
    \includegraphics[width=.85\textwidth]{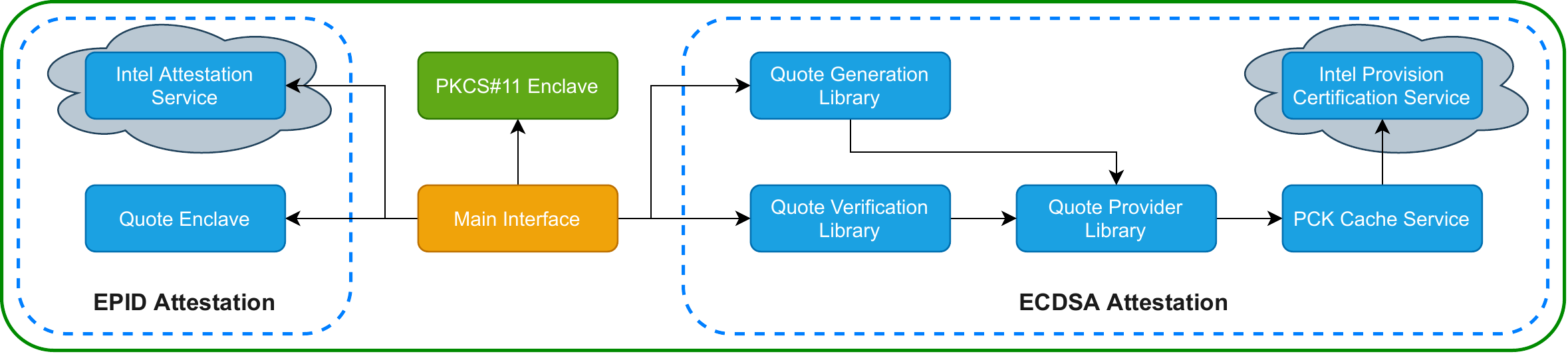}
    \caption{Architecture of \name\ supporting EPID and ECDSA attestations. The components in blue are trusted and built by Intel. The component in green is the trusted enclave implementing PKCS\#11. The component in orange is the main interface to integrate and control \name; it is untrusted (i.e., it does not hold any secret).} 
    \label{architect}
\end{figure*}
The foundation of ECDSA attestation infrastructure relies on the \emph{Provisioning Certification Enclave} (PCE), which serves as a local CA for the QE running on the same platform. To generate an ECDSA quote:
\begin{enumerate}
    \item QE generates an ECDSA key pair as attestation keys.
    \item QE locally attests itself to the PCE.
    \item Upon successful local attestation, the PCE issues a certificate for the QE's attestation public key. The certificate is signed by a platform-specific Provision Certification Key (PCK). Furthermore, Intel issues certificates for all PCKs in genuine platforms and signs them with the private key of Intel's root certificate.
    \item The ECDSA signature of the resulting quote can then be verified given the complete certificate chain to Intel's root certificate.
\end{enumerate}
An organization can set up a cache server containing the PCK certificates for all its machines after initially obtaining them from Intel's PCK web service. Hence, nodes can verify quotes without sending external requests over the Internet. 
 
In both models of the remote attestation, the challenger checks the quote's fields before its verification. Most importantly, the quote's \emph{MRENCLAVE} field must match the expected enclave's measurement. Otherwise, a malicious enclave could have generated the quote that deliberately leaks its secrets. Additionally, the challenger inspects the \texttt{report\_data} field for data-bound to the attesting enclave (e.g., an ECDH public key or a public key of a certificate's private key)

\subhead{Side-channel attacks on Intel SGX} Recently, several side-channel attacks on Intel SGX have been published, including the devastating Foreshadow attack~\cite{10.5555/3277203.3277277}. Foreshadow allows the attacker to extract all secrets and the attestation and sealing keys within an enclave. This attack breaks the entire TEE security assumptions in Intel SGX. Accordingly, Intel has released microcode patches and mitigation libraries to counter Foreshadow and other related attacks. However, we consider protection against side-channel attacks to be outside of the scope of this work. Furthermore, \name\ can be built using other TEE technologies as long as they provide remote attestation and sealing capabilities. 
 
\subsection{Threat Model and Assumptions}
\begin{enumerate}
    \item We assume that operators/malware can acquire the highest software privileges on a system (e.g., root privileges or even ring-0 on x86) through any traditional mechanisms (often via rootkits), including: known but unpatched vulnerabilities, zero-day vulnerabilities, and social engineering. Root-level access allows operators/malware to control devices and access the kernel.
    \item We assume systems utilize \name\ to protect private keys against exposure rather than misuse. In other words, \name\ acts as a secure vault of private keys. The adversary's sole objective is to expose private keys. Nonetheless, \name\ does not prevent \emph{misuse} by malicious operators (e.g., signing a maliciously crafted message). However, \name\ keeps track and logs requested operations in a rollback-resilient~\cite{203712} manner, which helps in \emph{detecting} misuse.
    \item We assume all CPUs have the latest microcode update and are implemented correctly by their manufacturers. Typically, users are motivated to choose CPUs that have no flaws. Unfortunately, \name\ does not guarantee its security when using poorly implemented CPUs vulnerable to side-channel attacks.
    \item In ECDSA attestation, we assume the involved organization (e.g., CA, CDN, or hosting provider) to have deployed its internal infrastructure securely and obtained PCK certificates for all nodes within its operation perimeter. Furthermore, each node has its corresponding PCK certificate authenticated by an internal CA within the organization. Therefore, it is a mandatory task to prevent a rogue/malicious node from authenticating itself as a benign node and request the provision of private keys from other nodes within the organization. 
\end{enumerate}

\section{\name\ Design}
This section presents the architecture of \name and shows how it works across websites, CAs and CDNs. Then, it illustrates the private key provision protocol that can securely transmit keys between nodes within the same organization. 

\subsection{Notation and APIs}

We utilize the following algorithms and APIs in the design of \name:
\begin{itemize}
    \item $q\gets \texttt{Quote}(x,t)$ generates a quote $q$, where $x$ denotes the value of \texttt{report\_data}, and $t$ is either \texttt{EPID} or \texttt{ECDSA}.
    \item $\texttt{VerifyQuote}(q,x,t)$ verifies a quote $q$ and checks that the field \texttt{report\_data} is equal to $x$ for attestation type $t$.
    \item $CSR\gets \texttt{GenCSR}()$ generates a key pair, fills and signs a $CSR$ by the generated private key, and returns the $CSR$ along with its public key $CSR.pk$.
 \item $Cert_{b}\gets \texttt{IssueCert}(CSR,Cert_{a}.sk)$ signs $CSR$ by the private key of $Cert_{a}.sk$, and returns a certificate $Cert_b$.
 \item $\texttt{VerifyCert}(Cert_b, Cert_a)$ verifies whether $Cert_b$ is issued by $Cert_a$.
 \item $(pk,sk)\gets \texttt{ECDH.KeyGen}()$ generates an ephemeral key pair.
 \item $k\gets \texttt{DeriveSharedKey}(pk_a, sk_b)$ derives a shared symmetric key based on ECDH public key $pk_a$ and private key $sk_b$.
 \item $c \gets \texttt{Encrypt}(m, k)$ and $m\gets \texttt{Decrypt}(c, k)$ are symmetric key algorithms.
 \item $\sigma \gets \texttt{Sign}(m, sk)$ and $\texttt{Verify}(\sigma, pk)$ are digital signature algorithms.
\end{itemize}
All verification APIs abort the running protocol in the event of a failure. Furthermore, in order to support auditability such that an admin can check the history of API calls and their parameters, \name\ utilizes a rollback-resilient protocol~\cite{203712} to log each invoked API.
\begin{figure*}
    \centering
    \includegraphics[width=.85\textwidth]{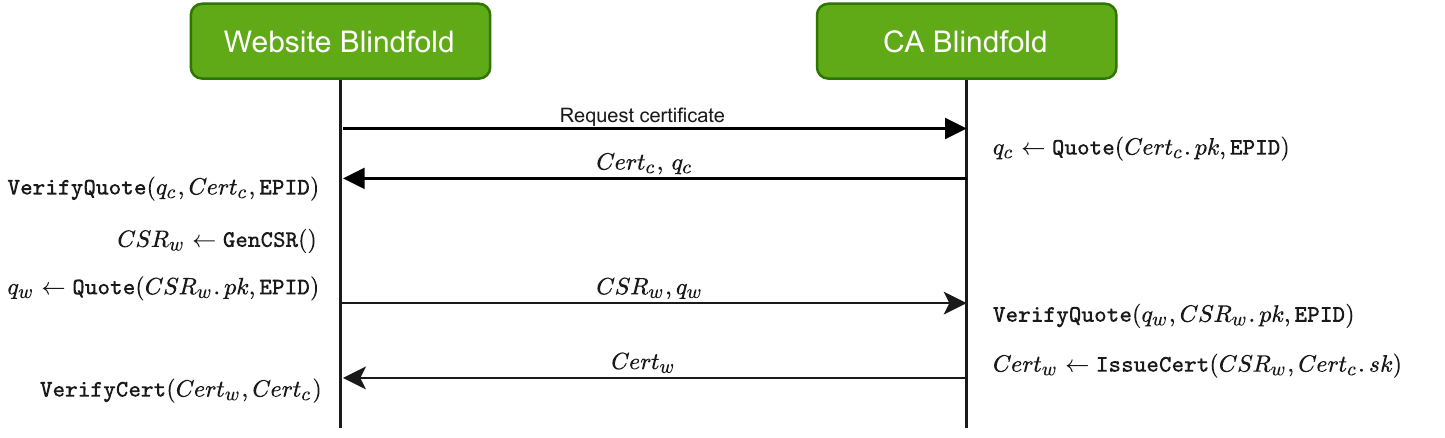}
    \caption{Interactions between a website and CA for certificate issuance during operational phase}
    \label{fig1}
\end{figure*} 
\subsection{Overview}
Ensuring the protection of private keys is a crucial objective, especially in the web context. We show how to achieve that objective besides having the ability to provision private keys between two \name\ instances securely. To this end, we utilize TEE technology with remote attestation and sealing features. In particular, we design \name\ by leveraging Intel SGX enclave.  Furthermore, the enclave implements PKCS\#11 interface, thereby, compatible software can easily integrate \name. 

The design of \name\ allows it to support both attestation models EPID and ECDSA as shown in Fig. \ref{architect}. \name\ can generate and verify ECDSA quotes, EPID quotes, and IAS verification reports on EPID quotes. Furthermore, \name\ instances can perform mutual attestation for building a secure channel. Typically,  \name\ utilizes ECDSA attestation when the attester and challenger nodes belong to the same organization (e.g., nodes of CA or CDN); otherwise, \name\ employs EPID attestation.

\subsection{Certificate Issuance}
We consider both the CA and website utilizing \name in the certificate issuance. Practically, certificate issuance consists of two main tasks, domain validation; and CSR generation and signing. Validating domain ownership is beyond the scope of our paper, and the CA may use ACME protocol~\cite{10.1145/3319535.3363192} to process this task. The main focus of \name\ is to secure the generation and signing of a CSR. The workflow between the CA and website starts with a setup phase followed by an operation phase. 

\subhead{Setup phase}
In the setup phase, the CA requests its \name\ to generate a key pair that matches its certificate key policy (e.g., 4096 bits RSA). Then, it creates a certificate $Cert_c$ binding the generated public key $Cert_c.pk$ to the \emph{Subject} identity. Subsequently, \name\ self-signs $Cert_c$ using the generated private key $Cert_c.sk$.  It is worth mentioning that \name\ seals the private key to untrusted storage during key generation. Next, \name\ creates an EPID quote $q_c$ to attest to the key generation by the enclave by setting $Cert_c.pk$ in the \texttt{report\_data}.  Finally, the CA publishes $Cert.c$ and the $q_c$, which assures the utilization of \name\ to protect the private key $Cert_c.sk$ on an authentic Intel SGX platform. 

Ideally, \name\  generates the quote once for the lifetime of the CA certificate. Nonetheless, in the event of upgrading \name\ (e.g., to add new features, bug fixes) or updating the underlying SGX firmware (e.g., software/microcode updates to enhance security or fixing newly discovered vulnerabilities), the CA needs to regenerate $q_c$. In particular, the \texttt{report} structure underlying $q_c$ contains a field that shows the Software Version Number (SVN) and the Trusted Computing Base (TCB) version. Hence, one can verify whether the CA has generated $q_c$ on an up-to-date platform. 
\begin{figure*}[t!]
    \centering
    \includegraphics[width=.85\textwidth]{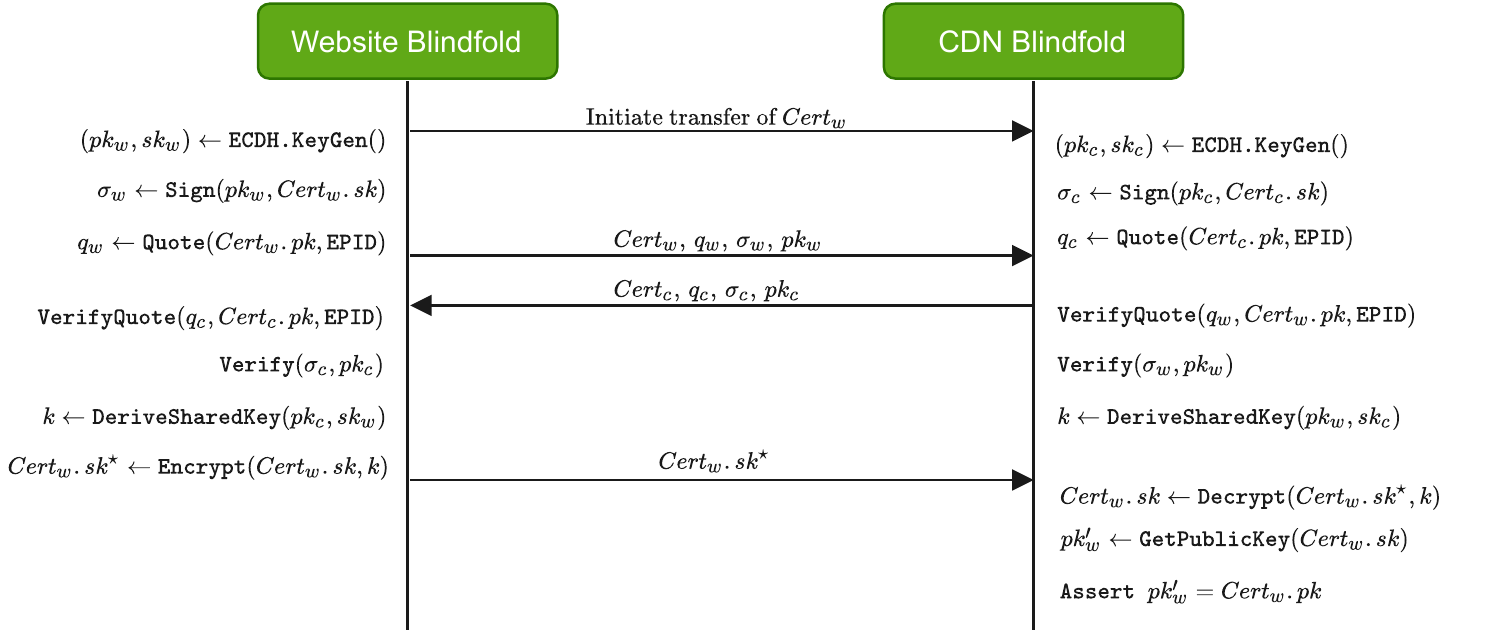}
    \caption{Transferring a website certificate's private key to a CDN}
    \label{fig2}
\end{figure*}

\subhead{Operation phase}
Once the CA is operational, websites can send certification requests to obtain domain certificates. Figure \ref{fig1} depicts the interactions between a CA and a website for certificate issuance. First, the website obtains the CA certificate $cert_c$, which contains the quote $q_c$ as a certificate extension. Typically $cert_c$ should be already installed as part of the operating system's trusted certificates store. Next, the website requests its \name\ to check $cert_c$ and $q_c$ for inconsistencies as follows: 
\begin{enumerate}
     \item The {\emph{MRENCLAVE}} value must match the correct measurement of \name's enclave. Otherwise, the CA could be using a modified version that may include unwanted features (e.g., deliberately exposing the enclave's secrets).
\item The \texttt{report\_data} field must contain the exact public key defined in the \texttt{Subject Public Key} field of $cert_c$. Otherwise, the CA could have generated $cert_c$ by tools other than  \name, implying that the private key for $cert_c$ has not been exclusively generated and managed by \name.
\end{enumerate}
Upon successful quote inspection, the website's \name\ communicates with IAS for verification since it is the only entity that can verify EPID quotes. If the verification passes successfully, \name\ generates a CSR $CSR_w$ with new key pair. Subsequently, \name\ generates an EPID quote $q_w$ where the \texttt{report\_data} field is set to the CSR's public key $CSR_w.pk$, thereby attesting to the generation of the key pair in the CSR. Next, the website forwards the $CSR_w$ and $q_w$ to the CA. Afterward, the CA's \name\ inspects and verifies $q_w$. Upon success, the CA creates a certificate $Cert_w$ and requests its \name\ to sign it by $Cert_c.sk$. Then, the CA returns the signed certificate  $Cert_w$ to the website. Finally, the website verifies that $Cert_c$ is the issuer of $Cert_w$ before accepting it. The web server must integrate with \name\ via PKCS\#11 interface to perform TLS handshakes and establish HTTPS sessions with its clients.

In general, CAs are not responsible for ensuring the protection of CSRs' private keys; however, we argue that doing so will enhance the security of the web PKI. Otherwise, one might question the point of issuing a certificate while not ensuring the protection of its private key. Hence, there is an incentive for CAs to do this task as it will reduce the number of possible revocations due to exposed keys. Moreover, CAs are central entities in that ecosystem; thereby, a healthy web PKI will positively impact their business model. It is an optional task in our design, and CAs might choose not to perform it.

\subsection{Integrating \name\ with a CDN}
The numbers of third parties, including CDNs and hosting providers that support requesting X.509 certificates on behalf of their customers, are increasingly growing~\cite{csp}. A CDN can utilize \name\ to interact with the CA on behalf of the website as depicted in Figure~\ref{fig1}. Once the CDN retrieves the certificate, it forwards both quotes (i.e., CA quote and CSR quote) to the website to attest to the protection of all involved private keys. On the other hand, the website may maintain total control over the certificate issuance process. Then, it can transfer the certificate's private key to the CDN's \name. Figure~\ref{fig2} shows the activity sequence for a website to move its certificate key securely to a CDN. The website and CDN utilize a modified SIGMA (SIGn-and-MAc) protocol to provide perfect forward secrecy via an authenticated Diffie-Hellman key exchange, where quotes serve as digital signatures.

We consider the website and CDN with certificates $Cert_w$ and $Cert_c$ to be utilizing \name\ to protect their private keys $Cert_w.sk$ and $Cert_c.sk$, respectively. This process starts with the website declaring its intention to send its certificate key to the CDN. Then, each entity individually utilizes its \name\ to generate ephemeral ECDH key pairs $(pk_w, sk_w)$ and $(pk_c, sk_c)$, respectively.  \name\ instances must mutually authenticate before establishing a secure channel to prevent MitM attacks. Accordingly, each instance signs its ephemeral public key $pk_w$, and $pk_c$ by the certificate's private key $Cert_w.sk$ and $Cert_c.sk$, respectively. Afterward, they exchange the certificates $Cert_w$ and $Cert_c$, the quotes $q_w$ and $q_c$, the signatures $\sigma_w$ and $\sigma_c$, and the ECDH public keys $pk_w$ and $pkc$. Subsequently, each instance verifies the received certificate's quote and verifies the signature on the ECDH public key generated by the certificate's private key. Upon success, both instances are now mutually authenticated and can establish a secure end-to-end channel. Accordingly, each instance derives a symmetric shared key $k$ based on its ECDH private key and the other instance's ECDH public key. Finally, the website's \name\ instance encrypts the $Cert_w.sk$ by $k$ and sends the ciphertext $Cert_w.sk^\star$ to the CDN. Finally, the CDN's \name\ instance decrypts $Cert_w.sk^\star$ using the same derived key $k$ and seal $Cert_w.sk$ to its storage. Certainly, the CDN server must integrate with its \name\ instance via PKCS\#11 to perform a TLS handshake and serve HTTPS content for that website.

\subsection{Provisioning of Private Keys}
Large organizations such as CDNs and CAs, to some extent, often have a cluster of servers across the globe. Mainly to improve web performance and user experience by reducing round trip latency. A typical CDN will have thousands of \name\ instances running across its nodes. Therefore, the domain certificate's private key must also exist in their \name\ instances to serve the web content for a given domain from these nodes. Hence, it is a mandatory requirement that \name\ supports the provision of private keys to other instances. 

\par The provision protocol simply involves two instances of \name\ running within the same organization. Hence, utilizing ECDSA attestation is far more convenient compared to EPID attestation. Initially, the organization establishes the internal infrastructure: a PCK cache server and an internal CA. The PCK cache server contains a list of PCK certificates for each Intel SGX platform in the organization. Furthermore, this list needs to be authenticated by the organization to guard against the injection of malicious nodes (see Section \ref{secAnalysis}-(d)). Hence, the organization generates a master key-pair $(mpk, msk)$ by a \name\ instance explicitly maintained by the admin. Then, the admin signs each PCK certificate in the cache server using $msk$, and install the $mpk$ on every node. Now that the infrastructure is ready, nodes can provision private keys. 

\begin{algorithm}
\DontPrintSemicolon
\KwIn{$NodeType$}
$(pk_a, sk_a) \gets \texttt{ECDH.KeyGen}()$\;
$q_a \gets \texttt{Quote}(pk_a, \texttt{ECDSA})$\;
\texttt{Send}$(pk_a, q_a)$\;
$(pk_b, q_b) \gets $\texttt{Receive}$()$\;
\texttt{Assert VerifyQuote}$(q_b, pk_b, \texttt{ECDSA})$\;
$k \gets \texttt{DeriveSharedKey}(pk_b, sk_a)$\;
\If{$NodeType$ = \texttt{SENDER}}{
	$c \gets \texttt{Encrypt}(\texttt{Certs}, k)$\;
	\texttt{Send}$(c)$\;
	}
	\Else {
	$c \gets \texttt{Receive}()$\;
	$\texttt{Certs} \gets \texttt{Decrypt}(c, k)$\;
	$\texttt{Store}(\texttt{Certs})$\;
	}
\caption{\texttt{ProvisionKeys}}
\label{alg:prov}
\end{algorithm}

Recall that in Fig.\ref{architect}, \name\ has access to the PCK cache service; hence, it can obtain the PCK certificates for any node within the organization. More importantly, \name\ verifies whether the obtained PCK certificates are signed by the organization's private key $msk$ before beginning this protocol. Besides generating ECDSA quotes, using PCK certificates allows both \name\ instances to perform mutual authentication. 

\par Within an organization such as a CDN and a CA, nodes run Algorithm~\ref{alg:prov} to provision certificates' private keys. A node can be either a sender or a recipient based on whether it has private keys to the certificates. The algorithm starts with generating an ECDH key-pair $(pk_a,sk_a)$. Afterward, it generates an ECDSA quote $q_a$ where the \texttt{report\_data} field is set to the ECDH public key $pk_a$. Subsequently, after exchanging the quotes and public keys, it verifies the received quote $q_b$ and checks whether the quote binds the exact received public key $pk_b$. Upon success, both nodes are now mutually authenticated and can build a secure end-to-end channel by deriving a shared symmetric key $k$. Finally, the certificates \texttt{Certs} private keys are encrypted by $k$ and sent to the recipient over the secure channel.

In practice, there is always a chance for a server hardware failure. Without an adequate backup and recovery procedure, the failure of the Intel SGX platform will result in a loss of certificates' private keys. One of the best security practices is to keep a backup of \name's state (i.e., sealed private keys) on a backup machine. The private keys within \name\ are non-extractable. Thereby, we cannot just utilize external backup destinations such as HSM.
\par Additionally, recall that the derived sealing key on Intel SGX is CPU-specific. Therefore, it is impossible to copy the sealed private keys to another platform simply. Instead, \name\ utilizes the provisioning protocol to transfer private keys to other \name\ instances for backup purposes. These instances can then be taken offline or operate in air-gap mode. Furthermore, the provisioning protocol allows \name\ to be portable as it can transfer its entire state (i.e., private keys) to other platforms within the same organization.

\section{Evaluation} 
In this section, we analyze the security of \name\, then we evaluate the performance for a prototype and report the results.
\subsection{Security Analysis}
\label{secAnalysis}
We analyze various possible attacks on \name\ and show how to counter them. 

\subhead{(a) Extracting private keys from \name}
Assuming an adversary has access to the platform of a target \name\ instance, the adversary's goal is to extract the private key. Note that the adversary can also be a rogue operator. The severity of a successful attack depends on the type of target platform, whether it belongs to a CA, a CDN, or a website. In \name, the private key is generated exclusively during run-time within the enclave. Then, it is sealed by the platform-specific sealing key, which is accessible only by \name's enclave. In other words, as long as the adversary neither has access to the enclave's protected memory nor the sealing key, then the adversary cannot extract the private key. 

One of the \name's assumptions is the correct implementation of CPUs TEE features; however, Intel SGX technology is susceptible to side-channel attacks in practice. Hence, we propose a solution to \emph{complicate} the attack against \emph{external} adversaries. \name\ can utilize threshold digital signature schemes for RSA and ECDSA~\cite{10.1007/3-540-48329-2_35,10.1145/3243734.3243859} to perform the private key generation and signing operations. In particular, rather than having a single \name\ enclave that generates and protects the private key, \name\ can employ $n$ enclaves running on different nodes within the same organization like a CA or a CDN to perform a distributed key generation. Then, each enclave will have a private key share such that at least $t$ instances can cooperate to perform the signing operation. In this case, the adversary must compromise and subvert at least $t$ different platforms, which should be more difficult to achieve than a single target platform. However, it is worth noting that threshold signing will add significant overhead, at minimum $t$ times the cost of a single sign operation, and the network delay associated with communicating shares of signature between participating enclaves.

\subhead{(b) Malicious access to \name} System software can communicate with \name\ instance via PKCS\#11 interface. Hence if the adversary has access to the target platform, it can interact with \name\ and send a \texttt{Sign} request to issue a bogus certificate or impersonate a web server, in the case of a CA or CDN platform, respectively. Fortunately, the PKCS\#11 standard requires authentication via PIN codes before performing the requested operation. However, this countermeasure is insufficient to deter a rogue operator or an adversary who knows a valid authentication code. Note that there is no solution to prevent misuse by malicious operators completely. For example, a rogue admin in a CA can misuse the private key to issue fraudulent certificates. Therefore, only auditing access logs such as \textit{Certificate Transparency Logs} can detect such misbehavior. Blindfold follows the same approach to detect misuse by operators. More specifically, \name\ maintains a log of all requested APIs which can be scanned frequently by the origin website to detect inconsistencies between TLS sessions logged at the CDN compared to logs at the web server. Moreover, to protect against a rollback attack of log state, \name\ utilizes rollback countermeasures~\cite{10.1145/3052973.3053034, 203712} to detect old states. Hence, the adversary's attack will not go undetected.

\subhead{(c) Intercepting private keys during provisioning to a remote node}
The ability to provision private keys to remote nodes is crucial to make \name\ practical to use within organizations. Furthermore, it allows \name\ to securely transfer private keys to remote nodes for backup and load balance. An adversary monitoring the traffic between the two nodes can launch a MitM attack to intercept the private keys during transit. However, in the \name\ design, both nodes perform the provisioning only via \name\ instances, only after successfully attesting their enclaves using ECDSA attestation quotes. Each quote contains an ECDH public key where the attesting enclave controls the corresponding ECDH private key. Hence, the adversary cannot replace the ECDH public key of one instance with another one she generated without invalidating the ECDSA signature. Consequently, upon successful ECDSA quote verification, both enclaves are now mutually authenticated and build a secure channel by deriving a shared secret key to encrypt the private keys for certificates in transit.

\subhead{(d) Installing a rogue \name\ instance within organization} The adversary may try to run a \name\ instance on its platform and request provisioning of the private keys. The provisioning in \name\ utilizes ECDSA attestation, which relies on an internal infrastructure built by the target organization. The infrastructure includes the PCK certificate cache server, which contains the PCK certificates for eligible nodes within the organization. Thus, the organization's nodes cannot authenticate the adversary's \name\ quote, thereby aborting the provisioning process. 

The adversary must compromise the PCK certificate cache server for this attack to succeed. Then, the adversary can inject her platform's PCK certificate into the cache server. Although this attack should be easy to detect, it is also preventable. First, the organization must deploy an internal CA, which signs each eligible PCK certificate on the cache server. Additionally, \name\ must successfully verify the counterparty's PCK certificate before initiating the provisioning protocol. As long as the internal CA private key is not compromised, the adversary cannot inject her PCK certificate successfully. Accordingly, \name\ instances will reject the adversary's request to provision private keys.

\subsection{Prototype Development}
To evaluate \name\ performance, we build a prototype based on a fork of Intel's repository for Crypto API Toolkit. The prototype implements most of the API defined in the standard PKCS\#11. In other words, it is a rendition of HSM that is built on Intel SGX, using SGX SDK  2.13 for Linux and SGX DCAP driver 1.10. The enclave exposes 71 ecalls and four ocalls. In addition, it contains two main components: (i) an enclave that implements the cryptographic operations and (ii) an untrusted shared library that acts as the main proxy for loading the enclave and redirecting calls with external processes. Essentially, the untrusted component is built as a shared library similar to HSM software engines so that external processes can load it.

We adjust the code in the prototype to meet the objectives of \name. In addition, we implement additional features that are not part of the PKCS\#11 interface. In particular, we add the ability to generate both ECDSA and EPID remote attestations for two purposes. First, it allows \name\ to attest to the exclusive generation of private keys by the enclave on an authentic Intel SGX platform. Second, remote attestation is a mandatory step to transfer private keys to remote \name\ instances securely.

\subsection{Interoperability Evaluation}
One of the main advantages of \name\ is that it adheres to the PKCS\#11 standard interface. Therefore, existing CA systems such as Boulder and web servers such as Apache and NGINX can integrate the prototype. We opted for Boulder, an open-source ACME CA implementation, to evaluate the interoperability between \name\ and other systems. Since Boulder already supports integration with HSM tokens via PKCS\#11, we can easily integrate \name\ by simply changing three lines of configuration in Boulder. 

Similarly, modern web servers such as NGINX and Apache utilize the OpenSSL library to perform cryptographic operations. OpenSSL can delegate cryptographic operations to PKCS\#11 compliant cryptography engines. By configuring OpenSSL to use \name\ as an HSM engine, NGINX and Apache can indirectly use \name\ to generate certificate key pair, perform TLS handshake, and establish HTTPS sessions. It is worth mentioning that utilizing \name\ alone by CDNs to establish TLS connection reduces the effectiveness of their firewall service, mainly due to how the firewall scans plaintext requests to filter out targeted attacks. Nonetheless, existing protocols that solve this problem~\cite{247664} can easily be integrated with \name. In other words, \name\ provides a secure key store in the context of a CDN similar to the Keyless SSL approach without its limitations, such as performing an extra round-trip to the origin key server before establishing TLS sessions.
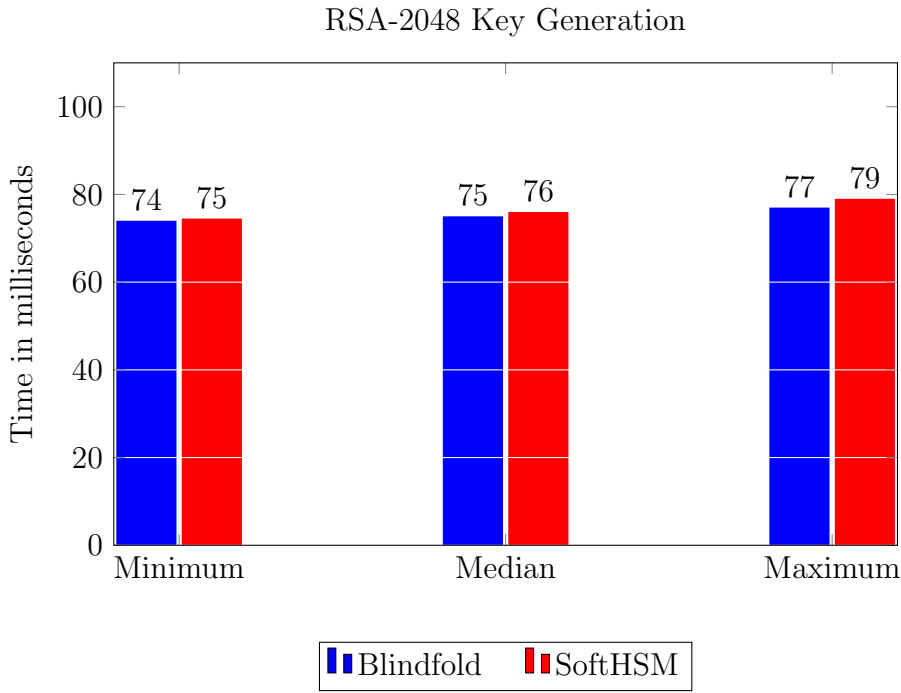
\begin{figure*}[htb!]
\centering
\begin{tikzpicture}
  \centering
  \begin{axis}[
        ybar, axis on top,
        title={RSA-2048 Key Generation},
        height=8cm, width=12cm,
        bar width=0.8cm,
        ymajorgrids, tick align=inside,
        major grid style={draw=white},
        enlarge y limits={value=.1,upper},
        ymin=0, ymax=100,
        legend style={
            at={(0.5,-0.2)},
            anchor=north,
            legend columns=-1,
            /tikz/every even column/.append style={column sep=0.5cm}
        },
        ylabel={Time in milliseconds},
        symbolic x coords={
           Minimum, Median, Maximum},
       xtick=data,
       nodes near coords={
        \pgfmathprintnumber[precision=0]{\pgfplotspointmeta}
       }
    ]
    \addplot [draw=none, fill=blue] coordinates {
      (Minimum,74)
      (Median, 75) 
      (Maximum,77) };
   \addplot [draw=none,fill=red] coordinates {
      (Minimum,74.5)
      (Median, 76) 
      (Maximum,79) };
    \legend{Blindfold, SoftHSM}
  \end{axis}
  \end{tikzpicture}
  \caption{Time measurements of RSA-2048 key generation by \name\ and SoftHSM}
\label{fig:perfKeyGen}
\end{figure*}

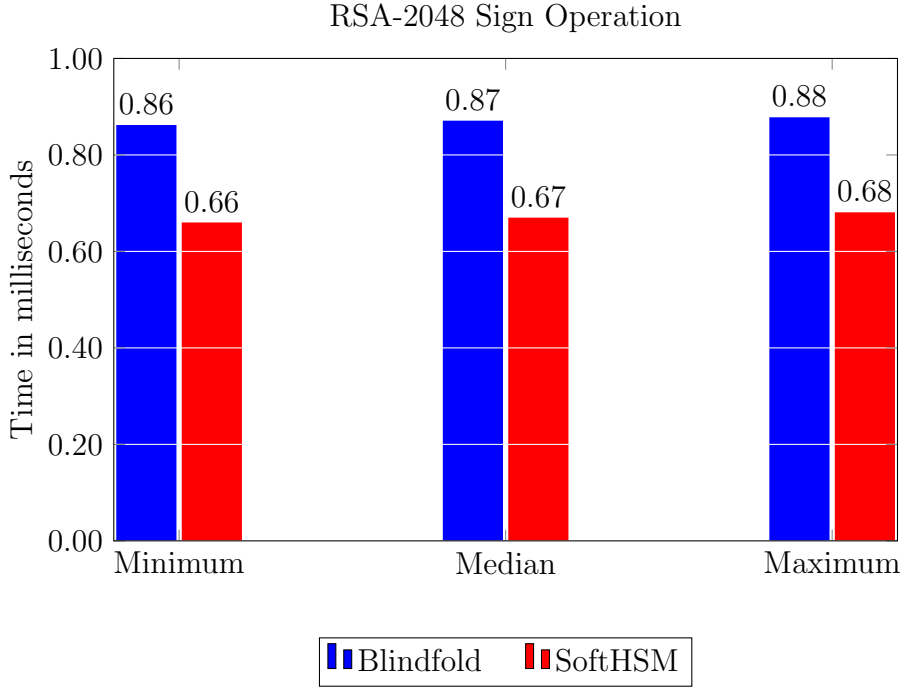
\begin{figure*}[htb!]
\centering
\begin{tikzpicture}
  \centering
  \begin{axis}[
        ybar, axis on top,
        title={RSA-2048 Sign Operation},
        height=8cm, width=12cm,
        bar width=.8cm,
        ymajorgrids, tick align=inside,
        major grid style={draw=white},
        ymin=0, ymax=1,
        legend style={
            at={(0.5,-0.2)},
            anchor=north,
            legend columns=-1,
            /tikz/every even column/.append style={column sep=0.5cm}
        },
        ylabel={Time in milliseconds},
        symbolic x coords={
           Minimum, Median, Maximum},
       xtick=data,
       nodes near coords={
        \pgfmathprintnumber[precision=2]{\pgfplotspointmeta}
       },
       yticklabel style={
            /pgf/number format/fixed,
            /pgf/number format/precision=2,
            /pgf/number format/fixed zerofill
        },
        scaled y ticks=false,
    ]
    \addplot [draw=none, fill=blue] coordinates {
      (Minimum, 0.862)
      (Median, 0.871) 
      (Maximum, 0.878) };
   \addplot [draw=none,fill=red] coordinates {
      (Minimum, .66)
      (Median, .67) 
      (Maximum, .681) };
    \legend{Blindfold, SoftHSM}
  \end{axis}
  \end{tikzpicture}
    \caption{Time measurements of RSA-2048 sign operation by \name\ and SoftHSM}
\label{fig:perfSign}
\end{figure*}

\begin{figure*}[htb!]
\centering
\begin{tikzpicture}
  \centering
  \begin{axis}[
        ybar, axis on top,
        title={Certificate Issuance},
        height=8cm, width=12cm,
        bar width=0.8cm,
        ymajorgrids, tick align=inside,
        major grid style={draw=white},
        enlarge y limits={value=.1,upper},
        ymin=0, ymax=5,
        legend style={
            at={(0.5,-0.2)},
            anchor=north,
            legend columns=-1,
            /tikz/every even column/.append style={column sep=0.5cm}
        },
        ylabel={Time in seconds},
        symbolic x coords={
           Minimum, Median, Maximum},
       xtick=data,
       nodes near coords={
        \pgfmathprintnumber[precision=2]{\pgfplotspointmeta}
       }
    ]
    \addplot [draw=none, fill=blue] coordinates {
      (Minimum,4.210)
      (Median, 4.273) 
      (Maximum,4.380) };
   \addplot [draw=none,fill=red] coordinates {
      (Minimum, 4.120)
      (Median, 4.235) 
      (Maximum, 4.520) };
    \legend{Blindfold, SoftHSM}
  \end{axis}
  \end{tikzpicture}
    \caption{Time measurements of certificate issuance from Boulder integrated with \name\ and SoftHSM}
\label{fig:perfIssuance}
\end{figure*}
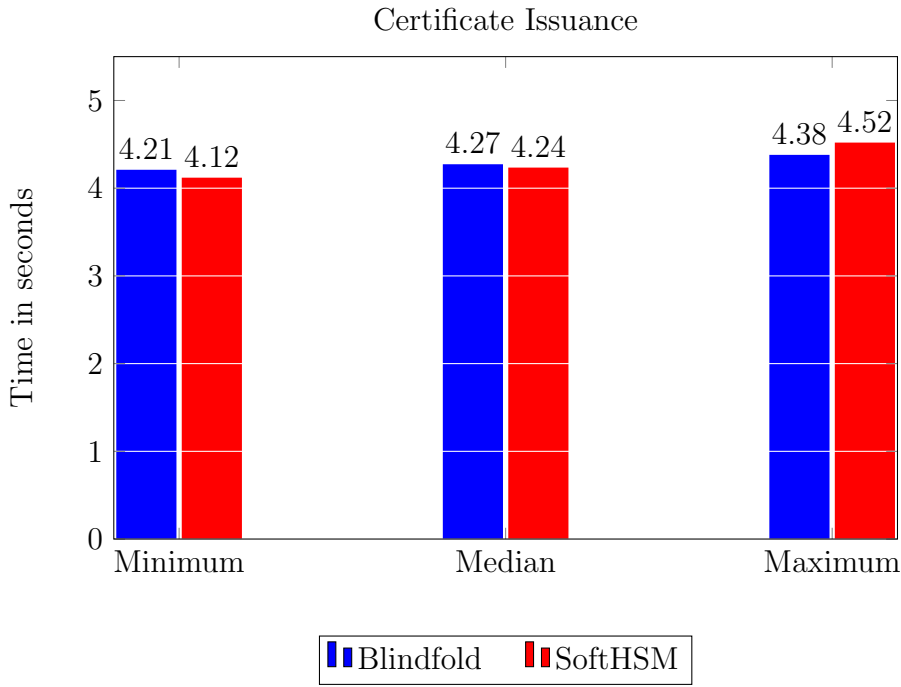

\subsection{Performance Evaluation}    
We carried out several experiments to measure the performance overhead of utilizing Intel SGX enclave by \name. In particular, we estimate the elapsed time for running cryptographic operations by an Intel SGX enclave. Key generation and signing are the most frequent operations invoked in \name\ regardless of the integrating software (i.e., CA, CDN, or web server). Then, we repeat the same experiments by running them natively on the normal untrusted CPU mode. Essentially, Intel SGX adds performance overhead, especially during the transition between the enclave and untrusted modes. For instance, each enclave transition imposes a cost of 8,400 CPU cycles, which is six times more costly than a typical system call~\cite{aublin2017talos}.  

Figures~\ref{fig:perfKeyGen} and ~\ref{fig:perfSign}  show the time measurement for invoking the key generation and signing operations in two different modes (i.e., with and without Intel SGX) for 1000 runs. The testing environment is a workstation with Ubuntu 20.04 OS, Intel Core i7-10750H CPU, and 16GB RAM. To this end, we run the experiments two times using \name\ and SoftHSM, respectively. Both offer a PKCS\#11 interface; however, SoftHSM provides the native CPU performance in the normal untrusted mode. One may argue that the comparison should be against a typical HSM; however,  comparing general-purpose CPUs against HSMs from a performance perspective is unfair and cannot correctly measure the performance overhead.
\par Interestingly, key generation time measurements are roughly close between the two modes even with the added enclave transition overhead. On the other hand, signing operation time measurements in \name\ lag behind SoftHSM by 30\%. However,  the security advantages of utilizing \name\ possibly far outweigh the performance overhead. One way to reduce the signing operation time is to batch the sign API calls to the enclave. A single enclave transition (i.e., invoking the sign ecall) signs $n$ certificate instead of just one. However, such an approach is not compliant with PKCS\#11 and will break interoperability with existing software.

We carried out an experiment to determine the time measurements to issue a certificate by a CA and the impact on user experience. Hence, we deploy two instances of Boulder; one is integrated with SoftHSM and the other one with \name. Then, we deploy Certbot, an ACME client, on the Amazon AWS EC2 t2.xlarge	node running Ubuntu 20.04 OS on Intel XEON processor and 16 GB RAM. We configure Certbot's IP settings to send requests to the deployed Boulder instances rather than the real Let's Encrypt server. Furthermore, we configure both Boulder instances with a fake DNS server to pass the domain validation checks such that it returns the Certbot node's IP address for any domain name. Then, we execute a script for Certbot to request 1000 certificates from both Boulder instances. Figure~\ref{fig:perfIssuance} shows the time measurement to issue a certificate from Boulder instances integrated with SoftHSM and \name, respectively. From a macro perspective, the time measurements for certificate issuance are roughly equal with a $0.01\%$ difference only. Hence, the performance overhead of utilizing Intel SGX has a negligible impact on the user (i.e., the domain owner) experience of using a \name-compliant CA.

\section{Conclusions}
\label{sec:conclusion}
There is no doubt that private keys protection against exposure is of utmost importance to ensure the security of any public-key protocol. In this paper, we designed \name, a generic approach that can protect private keys in any PKI-based system using a trusted execution environment. We implement a prototype to protect private keys in web PKI and solve the private keys sharing problem between websites and CDNs that is very common in practice and violates the basic assumption of public-key cryptography. Furthermore, \name\ also ensures public verifiability of CA private keys protection via remote attestation. We carried out several experiments to measure the performance overhead. We believe that the security benefits outweigh its performance overhead based on the experimental results.


\begin{thebibliography}{10}
\expandafter\ifx\csname url\endcsname\relax
  \def\url#1{\texttt{#1}}\fi
\expandafter\ifx\csname urlprefix\endcsname\relax\def\urlprefix{URL }\fi
\expandafter\ifx\csname href\endcsname\relax
  \def\href#1#2{#2} \def\path#1{#1}\fi

\bibitem{5772960}
R.~Langner, Stuxnet: Dissecting a cyberwarfare weapon, IEEE Security Privacy 9
  (2011) 49--51.

\bibitem{codecov}
Hashicorp,
  \href{https://discuss.hashicorp.com/t/hcsec-2021-12-codecov-security-event-and-hashicorp-gpg-key-exposure/23512}{Codecov
  security event and hashicorp gpg key exposure}.
\newline\urlprefix\url{https://discuss.hashicorp.com/t/hcsec-2021-12-codecov-security-event-and-hashicorp-gpg-key-exposure/23512}

\bibitem{prins2011diginotar}
J.~Prins, Diginotar certificate authority breach “operation black tulip”,
  Fox-IT (2011) 18.

\bibitem{ze}
K.~Zetter, Diginotar files for bankruptcy in wake of devastating hack, Wired
  magazine, September 1099 (2011).

\bibitem{symantec}
J.~Leyden,
  \href{https://www.theregister.com/2018/03/01/trustico_digicert_symantec_spat/}{23,000
  symantec certificates revoked following leak of private keys} (2018).
\newline\urlprefix\url{https://www.theregister.com/2018/03/01/trustico_digicert_symantec_spat/}

\bibitem{cryptoHacks2}
Selfkey, \href{https://selfkey.org/list-of-cryptocurrency-exchange-hacks/}{A
  comprehensive list of cryptocurrency exchange hacks}.
\newline\urlprefix\url{https://selfkey.org/list-of-cryptocurrency-exchange-hacks/}

\bibitem{cryptoHacks1}
P.~Thompson,
  \href{https://cointelegraph.com/news/most-significant-hacks-of-2019-new-record-of-twelve-in-one-year}{Most
  significant hacks of 2019 — new record of twelve in one year} (2020).
\newline\urlprefix\url{https://cointelegraph.com/news/most-significant-hacks-of-2019-new-record-of-twelve-in-one-year}

\bibitem{falliere2011w32}
N.~Falliere, L.~Murchu, E.~Chien, W32. stuxnet dossier, White paper, Symantec
  Corp., Security Response 5 (2011) 29.

\bibitem{6956557}
J.~Liang, J.~Jiang, H.~Duan, K.~Li, T.~Wan, J.~Wu, When https meets cdn: A case
  of authentication in delegated service, in: 2014 IEEE Symposium on Security
  and Privacy, 2014, pp. 67--82.

\bibitem{10.1145/2976749.2978301}
F.~Cangialosi, T.~Chung, D.~Choffnes, D.~Levin, B.~Maggs, A.~Mislove,
  C.~Wilson, Measurement and analysis of private key sharing in the https
  ecosystem, in: Proceedings of the 2016 ACM SIGSAC Conference on Computer and
  Communications Security, Association for Computing Machinery, Hofburg Palace,
  Vienna, Austria, 2016, p. 628–640.

\bibitem{10.1145/3127479.3127482}
C.~Wei, J.~Li, W.~Li, P.~Yu, H.~Guan, {STYX}: A trusted and accelerated
  hierarchical ssl key management and distribution system for cloud based cdn
  application, in: Proceedings of the 2017 Symposium on Cloud Computing, Santa
  Clara, California, 2017, p. 201–213.

\bibitem{247664}
S.~Herwig, C.~Garman, D.~Levin, Achieving keyless cdns with conclaves, in: 29th
  {USENIX} Security Symposium, 2020, pp. 735--751.

\bibitem{8567660}
R.~Ahmed, Z.~Zaheer, R.~Li, R.~Ricci, Harpocrates: Giving out your secrets and
  keeping them too, in: 2018 IEEE/ACM Symposium on Edge Computing (SEC),
  Bellevue, WA, USA, 2018, pp. 103--114.

\bibitem{9343007}
I.~Boureanu, D.~Migault, S.~Preda, H.~Alamedine, S.~Mishra, F.~Fieau,
  M.~Mannan, Lurk: Server-controlled tls delegation, in: IEEE 19th
  International Conference on Trust, Security and Privacy in Computing and
  Communications, Guangzhou, China, 2020, pp. 182--193.

\bibitem{boulder}
L.~Encrypt, \href{https://github.com/letsencrypt/boulder}{Boulder - an acme ca}
  (2015).
\newline\urlprefix\url{https://github.com/letsencrypt/boulder}

\bibitem{standard2020pkcs}
O.~Standard, {PKCS}\# 11 {Cryptographic Token Interface Base Specification
  Version} 3.0 (2020).

\bibitem{revoke}
T.~Spring,
  \href{https://threatpost.com/lets-encrypt-revoke-millions-tls-certs/153413/}{Let’s
  encrypt to revoke millions of tls certs} (2020).
\newline\urlprefix\url{https://threatpost.com/lets-encrypt-revoke-millions-tls-certs/153413/}

\bibitem{cab}
C.~Forum, \href{https://cabforum.org/baseline-requirements-documents/}{Baseline
  requirements documents (ssl/tls server certificates)}.
\newline\urlprefix\url{https://cabforum.org/baseline-requirements-documents/}

\bibitem{serrano2019complete}
N.~Serrano, H.~Hadan, L.~Camp, A complete study of pki (pki’s known
  incidents), SSRN (2019).

\bibitem{meli2019bad}
M.~Meli, M.~R. McNiece, B.~Reaves, How bad can it git? characterizing secret
  leakage in public github repositories., in: NDSS, 2019.

\bibitem{source}
D.~Sullivan,
  \href{https://searchcloudsecurity.techtarget.com/answer/Encryption-key-management-AWS-encryption-keys-got-exposed-now-what}{Our
  aws encryption keys were exposed accidentally, now what?} (2015).
\newline\urlprefix\url{https://searchcloudsecurity.techtarget.com/answer/Encryption-key-management-AWS-encryption-keys-got-exposed-now-what}

\bibitem{dlink}
M.~Mimoso,
  \href{https://threatpost.com/d-link-accidentally-leaks-private-code-signing-keys/114727/}{D-link
  accidentally leaks private code-signing keys} (2015).
\newline\urlprefix\url{https://threatpost.com/d-link-accidentally-leaks-private-code-signing-keys/114727/}

\bibitem{adobe}
J.~Nurminen, \href{https://twitter.com/jupenur/status/911286403434246144}{Adobe
  security team accidentally posts private pgp key on blog} (2017).
\newline\urlprefix\url{https://twitter.com/jupenur/status/911286403434246144}

\bibitem{kim2017certified}
D.~Kim, B.~J. Kwon, T.~Dumitra{\c{s}}, Certified malware: Measuring breaches of
  trust in the windows code-signing pki, in: Proceedings of the 2017 ACM SIGSAC
  Conference on Computer and Communications Security, 2017, pp. 1435--1448.

\bibitem{10.1145/2663716.2663755}
Z.~Durumeric, F.~Li, J.~Kasten, J.~Amann, J.~Beekman, M.~Payer, N.~Weaver,
  D.~Adrian, V.~Paxson, M.~Bailey, J.~Halderman, The matter of heartbleed, in:
  Proceedings of the 2014 Conference on Internet Measurement Conference,
  Vancouver, BC, Canada, 2014, p. 475–488.

\bibitem{zhang2014analysis}
L.~Zhang, D.~Choffnes, D.~Levin, T.~Dumitra{\c{s}}, A.~Mislove, A.~Schulman,
  C.~Wilson, Analysis of ssl certificate reissues and revocations in the wake
  of heartbleed, in: Proceedings of the 2014 Conference on Internet Measurement
  Conference, 2014, pp. 489--502.

\bibitem{fort}
Fortanix,
  \href{https://www.fortanix.com/products/data-security-manager/sdkms/}{Self-defending
  key management service}.
\newline\urlprefix\url{https://www.fortanix.com/products/data-security-manager/sdkms/}

\bibitem{awshsm}
Amazon, \href{https://aws.amazon.com/cloudhsm/}{Aws cloudhsm: Managed hardware
  security module (hsm) on the aws cloud}.
\newline\urlprefix\url{https://aws.amazon.com/cloudhsm/}

\bibitem{ibmhsm}
IBM, \href{https://www.ibm.com/cloud/hardware-security-module}{Ibm cloud
  hardware security module}.
\newline\urlprefix\url{https://www.ibm.com/cloud/hardware-security-module}

\bibitem{keyless}
CloudFlare, \href{https://www.cloudflare.com/en-ca/ssl/keyless-ssl/}{Overview
  of keyless ssl}.
\newline\urlprefix\url{https://www.cloudflare.com/en-ca/ssl/keyless-ssl/}

\bibitem{myers1999internet}
M.~Myers, C.~Adams, D.~Solo, D.~Kemp, Internet x. 509 certificate request
  message format, Request for Comments 2511 (1999).

\bibitem{anati2013innovative}
I.~Anati, S.~Gueron, S.~Johnson, V.~Scarlata, Innovative technology for cpu
  based attestation and sealing, in: Proceedings of the 2nd international
  workshop on hardware and architectural support for security and privacy,
  Vol.~13, 2013.

\bibitem{johnson2016intel}
S.~Johnson, V.~Scarlata, C.~Rozas, E.~Brickell, F.~Mckeen,
  Intel{\textregistered} software guard extensions: {EPID} provisioning and
  attestation services, White Paper 1 (2016) 119.

\bibitem{scarlata2018supporting}
V.~Scarlata, S.~Johnson, J.~Beaney, P.~Zmijewski, Supporting third party
  attestation for intel{\textregistered} {SGX} with intel{\textregistered} data
  center attestation primitives, Intel White paper (2018).

\bibitem{10.5555/3277203.3277277}
J.~Van~Bulck, M.~Minkin, O.~Weisse, D.~Genkin, B.~Kasikci, F.~Piessens,
  M.~Silberstein, T.~Wenisch, Y.~Yarom, R.~Strackx, Foreshadow: Extracting the
  keys to the intel sgx kingdom with transient out-of-order execution, in:
  Proceedings of the 27th {USENIX} Conference on Security Symposium, Baltimore,
  MD, USA, 2018, p. 991–1008.

\bibitem{203712}
S.~Matetic, M.~Ahmed, K.~Kostiainen, A.~Dhar, D.~Sommer, A.~Gervais, A.~Juels,
  S.~Capkun, {ROTE}: Rollback protection for trusted execution, in: 26th
  {USENIX} Security Symposium, Vancouver, BC, 2017, pp. 1289--1306.

\bibitem{10.1145/3319535.3363192}
J.~Aas, R.~Barnes, B.~Case, Z.~Durumeric, P.~Eckersley, A.~Flores-L\'{o}pez,
  J.~A. Halderman, J.~Hoffman-Andrews, J.~Kasten, E.~Rescorla, S.~Schoen,
  B.~Warren, Let's encrypt: An automated certificate authority to encrypt the
  entire web, in: Proceedings of the 2019 ACM SIGSAC Conference on Computer and
  Communications Security, CCS '19, London, United Kingdom, 2019, p.
  2473–2487.

\bibitem{csp}
L.~Encrypt,
  \href{https://community.letsencrypt.org/t/web-hosting-who-support-lets-encrypt/6920}{Web
  hosting providers let's encrypt support status} (2020).
\newline\urlprefix\url{https://community.letsencrypt.org/t/web-hosting-who-support-lets-encrypt/6920}

\bibitem{10.1007/3-540-48329-2_35}
C.~Li, T.~Hwang, N.~Lee, Remark on the threshold rsa signature scheme, in:
  Advances in Cryptology, Santa Barbara, California, USA, 1994, pp. 413--419.

\bibitem{10.1145/3243734.3243859}
R.~Gennaro, S.~Goldfeder, Fast multiparty threshold ecdsa with fast trustless
  setup, in: Proceedings of the 2018 ACM SIGSAC Conference on Computer and
  Communications Security, Toronto, Canada, 2018, p. 1179–1194.

\bibitem{10.1145/3052973.3053034}
V.~Karande, E.~Bauman, Z.~Lin, L.~Khan, {SGX}-log: Securing system logs with
  {SGX}, Abu Dhabi, United Arab Emirates, 2017.

\bibitem{aublin2017talos}
P.~Aublin, F.~Kelbert, D.~O’keeffe, D.~Muthukumaran, C.~Priebe, J.~Lind,
  R.~Krahn, C.~Fetzer, D.~Eyers, P.~Pietzuch, Talos: Secure and transparent tls
  termination inside {SGX} enclaves, Imperial College London, Tech. Rep 5
  (2017).

\end{thebibliography}
\end{document}